\documentclass[letterpaper,english]{IEEEtran}
\usepackage[T1]{fontenc}
\usepackage[latin9]{inputenc}
\usepackage{float}
\usepackage{amsthm}
\usepackage{amssymb}
\usepackage{graphicx}

\makeatletter

\providecommand{\tabularnewline}{\\}
\floatstyle{ruled}
\newfloat{algorithm}{tbp}{loa}
\providecommand{\algorithmname}{Algorithm}
\floatname{algorithm}{\protect\algorithmname}

\theoremstyle{plain}
\newtheorem{thm}{\protect\theoremname}
\theoremstyle{plain}
\newtheorem{lem}[thm]{\protect\lemmaname}

\usepackage{cite}

\usepackage{algorithmic}

\usepackage[cmex10]{amsmath}
\usepackage{url}

\makeatother

\usepackage{babel}
\providecommand{\lemmaname}{Lemma}
\providecommand{\theoremname}{Theorem}

\begin{document}

\title{Minimum-Variance Importance-Sampling Bernoulli Estimator for Fast
Simulation of Linear Block Codes over Binary Symmetric Channels}

\author{Gianmarco~Romano,~\IEEEmembership{Member,~IEEE}, and Domenico~Ciuonzo,~\IEEEmembership{Student~Member,~IEEE}%
\thanks{The authors are with the Department of Industrial and Information Engineering, Second University of Naples, via Roma, 29, 81031 Aversa (CE), Italy. Email: \{gianmarco.romano, domenico.ciuonzo\}@unina2.it}}
\maketitle
\begin{abstract}
In this paper the choice of the Bernoulli distribution as biased distribution
for importance sampling (IS) Monte-Carlo (MC) simulation of linear
block codes over binary symmetric channels (BSCs) is studied. Based
on the analytical derivation of the optimal IS Bernoulli distribution,
with explicit calculation of the variance of the corresponding IS
estimator, two novel algorithms for fast-simulation of linear block
codes are proposed. For sufficiently high signal-to-noise ratios (SNRs)
one of the proposed algorithm is SNR-invariant, i.e. the IS estimator
does not depend on the cross-over probability of the channel. Also,
the proposed algorithms are shown to be suitable for the estimation
of the error-correcting capability of the code and the decoder. Finally,
the effectiveness of the algorithms is confirmed through simulation
results in comparison to standard Monte Carlo method. \end{abstract}
\begin{IEEEkeywords}
Binary symmetric channel (BSC), importance sampling (IS), linear block
codes, Monte-Carlo simulation.
\end{IEEEkeywords}



\section{Introduction}

\label{sec:introduction}

\IEEEPARstart{T}{he} Monte-Carlo (MC) simulation is a general method
to estimate performances of complex systems for which analytical solutions
are not available or mathematically tractable and it is extensively
used in the analysis and design of communications systems\cite{Jeruchim,Tranter2004}.
The MC method has also been extensively employed to evaluate the performances
of forward-error-correcting (FEC) codes with different decoding algorithms,
in terms of probability of bit error (BER) or word error (WER), for
which, in many cases, is not possible to obtain exact closed-form
expressions\cite{Morelos-Zaragoza2006,Benedetto1999,Proakis2008}.
In general an upper bound is available for any linear block code,
however the error correcting capability of the code is required \cite{Benedetto1999,Proakis2008}.
The MC method is also used as verification tool in the design, development
and implementation of decoding algorithms.

The computational complexity of the MC method is given by the number
of generated random samples that are needed to obtain a reliable estimate
of the parameters of interest. In the case of FEC codes, estimation
of low BER or WER requires a high number of generated codewords to
obtain results of acceptable or given accuracy, thus leading to prohibitive
computational complexity. Furthermore, for very long codes the computational
complexity is high even for small number of generated words, since
the decoding complexity increases the simulation time considerably.
A practical case is represented by low-density parity-check (LDPC)
codes \cite{Gallager1962,MacKay1999,Lin2004}, for which it is crucial
to examine the performances at very low probability of error in order
to avoid error floors, i.e. the rate of decrease of the probability
of error is not as high as at lower SNRs (i.e. in the waterfall region)
\cite{Richardson2003,Chilappagari2006}. One of the impediments in
the adoption of LDPC codes in fiber-optics communications, where the
order of magnitude of the probability of error of interest is $10^{-12}$
and below, has been the inability to rule out the existence of such
floors via analysis or simulations \cite{Smith2010}. While for some
LDPC codes it is possible to predict such floors, in many other cases
the MC method is the only tool available. LDPC codes are also employed,
for example, in nanoscale memories \cite{Ghosh2011}, where a majority-logic
decoder is chosen instead of soft iterative decoders as these may
not be fast enough for error correction; therefore an efficient method
to estimate the performances of hard-decision decoding at very low
WERs is extremely desirable.

Several mathematical techniques have been proposed in the literature
in order to reduce the computational complexity of the MC method and
estimate low WERs with the same accuracy%
\footnote{One possibility to cope with the computational complexity of the MC
method is to adopt more powerful hardware in order to reduce the generation
and processing time of each codeword; this might constitute a practical
solution to reduce the overall simulation time. Nevertheless, the
increased system complexity requires more time per sample and compensates
the reduction of execution time, thus limiting the achievable gain.%
} \cite{Smith1997}. Importance sampling (IS) is regarded as one of
the most effective variance-reduction techniques and it is widely
adopted to speed up simulation of rare events, i.e. events that occur
with very low probability\cite{Rubinstein2008}. The idea is to increase
the frequency of occurrence of rare events, by means of a biased distribution.
The optimal biased IS distribution is known, but it cannot be used
in practice since it depends on the parameter to be estimated itself.
Therefore, a number of sub-optimal alternatives have been developed
in the literature \cite{Smith1997,Srinivasan2002}. Some of them are
obtained by restricting the search of the biased distribution to a
parametric family of simulation distributions; then the parameters
are derived as minimizers of the estimator variance or other related
metrics, such as the cross-entropy \cite{Rubinstein2008,Rubinstein2004}.
The choice of the family of biased distribution is somewhat arbitrary
and may depend on the specific application of the IS method \cite{Rubinstein2008}.
In the case of FEC, the rare event corresponds to the decoding error
and the IS method, in order to be effective, needs to generate more
frequently the codewords that are likely to be erroneously decoded.
The mathematical structure of the code, or some performance parameter
of the code, such as the minimum distance and/or the number of correctable
errors or, in the case of LDPCs, the minimum size of the absorbing
sets in their Tanner graphs, may be taken into account to choose a
good family \cite{Sadowsky1990,Xia2003}. In \cite{Mahadevan2007},
an SNR-invariant IS method is proposed, which, though independent
of the minimum distance of the code, provides better estimates when
the error-correcting capability of the decoder is available. In this
paper we consider generic linear block codes and we do not make any
assumption on specific parameter or structure of the code.

In this paper a specific problem is considered: (i) which is the best
joint independent Bernoulli distribution that can be used as biased
distribution for IS estimation of block linear code performances and
(ii) what are the strengths and limitations of this solution. The
choice of such family of distributions is arbitrary and it is motivated
by the fact that the random generator required for the IS method is
of the same type of that required in the standard MC method and hence
is made because of its simplicity rather than taking into account
the specific structure or properties of codes. On the other hand,
since the study is restricted to the parametric family of the joint
independent Bernoulli distributions, the gain in computational complexity
that is obtained is limited by this choice, as sub-optimal IS distributions
that lead to smaller IS estimator variance may exist.

Another performance measure for FEC codes is the minimum distance
of the code and/or the error correcting capability of the code or
decoder, i.e. the maximum number of errors that a specific couple
(code, decoder) are able to correct. The minimum distance of codes
can be estimated to overcome the computational complexity required
by the exhaustive search, which increases exponentially with the length
of the information word. In \cite{Berrou2002} the error impulse method
is proposed for linear codes and is based on the properties of the
error impulse response of the soft-in soft-out decoder and its error-correcting
capability. Due to the sub-optimality of the iterative decoder employed
with LDPC codes, the error impulse method can lead to wrong estimates
of minimum distance. For this class of codes the method has been improved
in \cite{Hu2004} and \cite{Daneshgaran2006}. More recently, integer
programming methods have been used to calculate either the true minimum
distance or an upper bound \cite{Punekar2010}. Alternatively, a branch-and-cut
algorithm for finding the minimum distance of linear block codes has
been proposed in \cite{Keha2010}. In this paper a novel MC method
to estimate the error correcting capability of the code/decoder is
derived.

Summarizing, the main contributions are the following: (i) analytical
derivation of the optimal importance sampling distribution among the
family of Bernoulli distributions, with explicit calculation of the
variance of the corresponding IS estimator and proof of convexity;
(ii) derivation of two algorithms for fast-simulation, one to estimate
numerically the optimal parameter of the importance sampling distribution
and one that is invariant to SNR; (iii) derivation of one algorithm
for efficiently estimate the number of correctable errors. Some illustrative
numerical examples of application of the proposed algorithms, for
BCH and LDPC codes, are also provided.

The proposed fast-simulation algorithms achieve large gains over standard
MC simulation for a vast variety of communication systems where linear
block codes are employed over binary symmetric channels (BSC). They
are simple to implement because they require only small modifications
to the standard MC method, as the same random sample generator can
be maintained and only the parameter of the Bernoulli generator is
changed. Furthermore, in most practical situations the SNR-invariant
version of the algorithm allows to efficiently obtain entire curves
of performance, e.g. WERs corresponding to various SNRs, \emph{by
just running one IS simulation at one sufficiently high SNR}. In such
a case the gain with respect to (w.r.t.) the standard MC simulation
is even higher, as the number of simulation runs is dramatically reduced
to one.

The outline of the paper is the following: in Sec. \ref{sec:system-model}
the system model is introduced and some preliminaries on MC and IS
method are given; the main results of the paper are presented in Sec.
\ref{sec:sub-optimal-importance-sampling}; in Sec. \ref{sec:algorithms}
fast-simulation algorithms are formulated and some examples are shown
in Sec. \ref{sec:examples}; finally, in Sec. \ref{sec:conclusions}
some concluding remarks are given; proofs are confined to the appendices.

\emph{Notation }- Lower-case bold letters denote vectors; the function
$wt\left(\mathbf{z}\right)$ returns the number of 1's in the binary
vector $\mathbf{z}$; $\mathbb{E}\left[\cdot\right]$ and \emph{$\mathrm{var}\left[\cdot\right]$}
denote expectation and variance operators, respectively; $\left\lceil \cdot\right\rceil $
denotes the ceiling operator; $P\left(\cdot\right)$ and $f\left(\cdot\right)$
are used to denote probabilities and probability mass function (pmf);
$\mathcal{B}\left(i,p\right)$ denotes the pmf of the $n$-dimensional
multivariate independent Bernoulli variable $\mathbf{z}$, with parameter
$p$, i.e. $f\left(\mathbf{z};p\right)=p^{i}\left(1-p\right)^{n-i}$,
where $i=wt\left(\mathbf{z}\right)$; $\mathbb{E}_{p}\left[\cdot\right]$
and \emph{$\mathrm{var}_{p}\left[\cdot\right]$} denote expectation
and variance operators with respect to the joint Bernoulli distribution
of parameter $p$, respectively; finally, the symbols $\sim$ and
$\oplus$ mean ``distributed as'' and ``modulo-2 addition'', respectively.

\section{System model}

\label{sec:system-model}

\begin{figure}
\begin{centering}
\includegraphics[width=1\columnwidth]{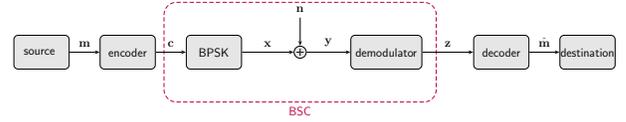}
\par\end{centering}

\caption{\label{fig:system-model}Illustrative block scheme of a communication
system.}
\end{figure}

A communication system where binary codewords are transmitted over
a BSC with transition probability $p$ is shown in Fig. \ref{fig:system-model}.
A codeword $\mathbf{c}$, belonging to the block code $\mathcal{C}\subset\mathcal{X}^{n}=\left\{ 0,1\right\} ^{n}$
is obtained by encoding message word $\mathbf{m}\in\mathcal{X}^{k}$;
at the output of the channel a word $\mathbf{z}\in\mathcal{X}^{n}$,
corrupted by noise, is observed. The decoder's task is to possibly
recover $\mathbf{m}$ given the observed $\mathbf{z}$. The BSC may
represent, for example, an additive white Gaussian noise (AWGN) channel
with binary phase-shift keying (BPSK) modulation and hard-decision
at the receiver, as shown in Fig. \ref{fig:system-model}. 

Performances of linear block codes over noisy channels are measured
by the probability of decoding error, i.e. the probability that a
decoded word is different from the transmitted message word, because
the block code was not able to correct the errors due to the channel.
This probability is also called probability of word error or WER.
This event occurs when the error pattern is not a co-set leader (under
the assumption that syndrome decoding is employed). Calculation of
WER is often very complex and some upper bounds are available \cite{Benedetto1999}. 

The WER, denoted as $P(e)$ hereinafter, can be expressed in terms
of an indicator function $I(\mathbf{z})$ that equals to 1 when the
received word is erroneously decoded and 0 otherwise; its explicit
form is given as 
\begin{equation}
P\left(e\right)=\sum_{\mathbf{z}\in\mathcal{X}^{n}}I\left(\mathbf{z}\right)f\left(\mathbf{z}\right)=\mathbb{E}_{p}\left[I\left(\mathbf{z}\right)\right].\label{eq:WER-MC}
\end{equation}
Note that the indicator function hides the specific decoding algorithm
employed. The effect of the BSC channel is to flip some bits, which
can be mathematically expressed by $\mathbf{z}=\mathbf{c}\oplus\mathbf{e}$,
with $\mathbf{e}\sim\mathcal{B}\left(wt\left(\mathbf{e}\right),p\right)$.
Since the code is linear and the channel symmetric, without loss of
generality (w.l.o.g.) the transmission of the codeword of all zeros
is assumed, i.e. $\mathbf{c}=\mathbf{0}$, and hence the output of
the channel $\mathbf{z}=\mathbf{e}$, i.e. equals the error pattern
$\mathbf{e}$.

\subsection{Monte-Carlo simulation}

\label{sub:monte-carlo-simulation}

In the MC simulation method the WER is estimated as follows 
\begin{equation}
\hat{P}_{MC}\left(e\right)=\frac{1}{N}\sum_{i=1}^{N}I\left(\mathbf{z}_{i}\right),\label{eq:mc-wer}
\end{equation}
where $\mathbf{z}_{i}$ are generated according the distribution of
the random variable $\mathbf{z}$. It is known that the MC estimator
(\ref{eq:mc-wer}) is unbiased and its variance 
\begin{equation}
\mathrm{var}\left[\hat{P}_{MC}\left(e\right)\right]=\frac{P\left(e\right)\left(1-P\left(e\right)\right)}{N}\label{eq:var-mc}
\end{equation}
is inversely proportional to $N$ (see, for example, \cite{Rubinstein2008}),
then it can be made arbitrarily small as $N$ grows, thus increasing
the accuracy of the estimator. Rather than studying the variance it
is often preferable to consider as accuracy of the estimator the relative
error \cite{Rubinstein2008}, defined as 
\begin{equation}
\kappa\triangleq\frac{\sqrt{\mathrm{var}\left[\hat{P}_{MC}\left(e\right)\right]}}{P\left(e\right)}.\label{eq:re}
\end{equation}
In standard MC simulation $\kappa$ becomes 
\begin{equation}
\kappa=\sqrt{\frac{1-P\left(e\right)}{P\left(e\right)N}},\label{eq:re-mc}
\end{equation}
and, for small probabilities of error ($P\left(e\right)\ll1$), it
is well approximated as 
\begin{equation}
\kappa\simeq\frac{1}{\sqrt{P\left(e\right)N}}.\label{eq:re-approx}
\end{equation}
It follows that the number of generated samples needed to achieve
a given $\kappa$ is 
\begin{equation}
N\simeq\frac{1}{\kappa^{2}P\left(e\right)}.\label{eq:Nsamples}
\end{equation}
Eq. \eqref{eq:Nsamples} shows that the number of samples needed to
obtain a given $\kappa$ is inversely proportional to $P\left(e\right)$
and becomes soon very high and often impractical as $P\left(e\right)$
decreases. For example with a relative error of $10\%$, at least
$N\simeq10^{2}/P\left(e\right)$ samples are needed to obtain the
desired accuracy.

\begin{algorithm}
\caption{\label{alg:standard-Monte-Carlo}Standard MC simulation algorithm}
\texttt{%
\begin{algorithmic}
\STATE{totWords = 0}
\STATE{WERre = 1}
\WHILE{ (WERre > re) and (totWords < maxNumWords)}
\STATE{ $\mathbf{z}$ = rand(n, numWords) < p } \COMMENT{ BSC output }
\STATE{ $\hat{\mathbf{m}}$ = decode($\mathbf{z}$) } \COMMENT{ decoder output }
\IF{ $wt(\hat{\mathbf{m}}) > 0 $ }
\STATE{ totWErr = totWErr + 1}
\ENDIF
\STATE{ totWords = totWords + numWords }
\IF{ totWords > minNumWords }
\STATE{ update WERre } \COMMENT{ relative error }
\ENDIF
\ENDWHILE
\end{algorithmic}
}
\end{algorithm}

Algorithm \ref{alg:standard-Monte-Carlo} represents a generic implementation
of MC simulation for estimation of $P\left(e\right)$ in BSCs \cite{Rubinstein2008,Jeruchim}.
The algorithm depends on three parameters: \texttt{re, minNumWords},
\texttt{maxNumWords}. The first parameter, \texttt{re}, is the relative
error and it is computed according to (\ref{eq:re-mc}) or its approximation
(\ref{eq:re-approx}), where $P\left(e\right)$ is replaced by $\hat{P}_{MC}\left(e\right)$,
i.e. the current estimate. The second parameter, \texttt{minNumWords,}
represents the minimum number of words needed to obtain a sufficiently
accurate estimate of $\kappa$. Once a confident estimate of $\kappa$
is obtained, a stop condition on the relative error can be employed.
In practice in most cases a relative error of $10\%$, i.e. $\kappa=0.1$,
may suffice, as often only the order of magnitude of the estimate
is of interest. Finally, \texttt{maxNumWords} represents the maximum
number of generated words and it is used to implement a second stop
condition that prevents the simulation to run too long.

Alternative stopping rules for MC simulations can also be considered.
One common rule consists of fixing the number of generated word before
running the simulation and the accuracy is estimated at the end of
simulation \cite{Jeruchim}. Another rule, analyzed in \cite{Mendo2006},
is based on the number of errors: when a given number has been reached,
then the simulation stops. The advantage of this second rule is that
it does not require to know the sample size and can achieve a given
accuracy.

\subsection{Importance sampling}

\label{sec:importance-samplig}

In IS simulation the WER is expressed by the following equivalent
of \eqref{eq:WER-MC} 
\begin{equation}
P\left(e\right)=\sum_{\mathbf{z}\in\mathcal{X}^{n}}I\left(\mathbf{z}\right)\frac{f\left(\mathbf{z}\right)}{f^{*}\left(\mathbf{z}\right)}f^{*}\left(\mathbf{z}\right),\label{eq:IS-wer}
\end{equation}
 where $f^{*}\left(\mathbf{z}\right)$ is a different pmf for which
the sum in \eqref{eq:IS-wer} exists. The corresponding estimator
is
\begin{eqnarray}
\hat{P}_{IS}\left(e\right) & = & \frac{1}{N}\sum_{i=1}^{N}I\left(\mathbf{z}_{i}\right)\frac{f\left(\mathbf{z}_{i}\right)}{f^{*}\left(\mathbf{z}_{i}\right)},\label{eq:is-pe}\\
 & = & \frac{1}{N}\sum_{i=1}^{N}I\left(\mathbf{z}_{i}\right)W\left(\mathbf{z}_{i}\right)\label{eq:is-estimator}
\end{eqnarray}
where $\mathbf{z}_{i}\sim f^{*}\left(\cdot\right)$ and the ratio
\begin{equation}
W\left(\mathbf{z}\right)\triangleq\frac{f\left(\mathbf{z}\right)}{f^{*}\left(\mathbf{z}\right)}
\end{equation}
is referred to as the likelihood ratio or weighting function. The
estimator in \eqref{eq:is-estimator} is called the \emph{IS estimator}
and is a generalization of the simple MC estimator in \eqref{eq:mc-wer},
that can be obtained as special case (i.e. $f^{*}\left(\mathbf{z}\right)=f\left(\mathbf{z}\right)$). 

The distribution $f^{*}\left(\mathbf{z}\right)$ is called the IS
or biased distribution and as long as the random generation of samples
is under our control, as in the case of MC simulation, it is possible
to choose any distribution. However, it is crucial to choose the IS
distribution such that the variance of the IS estimator is minimized.
The optimal distribution is known from theory (see for example \cite{Smith1997,Rubinstein2008})
and it is given by 
\begin{equation}
f_{opt}^{*}\left(\mathbf{z}\right)=\frac{I\left(\mathbf{z}\right)f\left(\mathbf{z}\right)}{P\left(e\right)}.
\end{equation}
This distribution leads to zero variance: this comes at no surprise
since $f_{opt}^{*}\left(\mathbf{z}\right)$ contains $P\left(e\right)$
(which is the true value of the parameter being estimated). For this
reason, the optimal solution cannot be used for MC simulation. Nonetheless,
significant gains in simulation time can be achieved with sub-optimal
biased distributions. Several methods to find sub-optimal biased distributions
have been developed and the interested reader can refer to the comprehensive
tutorial in \cite{Smith1997}. One important goal in searching a sub-optimal
IS distribution is to obtain a probability distribution from which
samples can be easily generated and that, at the same time, provides
a weighted estimator with as low variance as possible.

\section{Sub-optimal Importance Sampling}

\label{sec:sub-optimal-importance-sampling}

The main problem in the design of IS simulations is to find sub-optimal
distributions that lead to low variance of the IS estimator. The problem
can be simplified if the search is limited within a parametric family
of distributions, since the problem can be recast into a standard
optimization w.r.t. a finite number of parameters. Also, a proper
choice of the parametric family can reduce the computational complexity
due to the generation of random samples. In this paper \emph{the family
of Bernoulli distributions with parameter $q$ is considered}, thus
maintaining the simplicity of random generation of error patterns,
since no change of the random generator is required. In practice the
WER for a BSC with cross-over probability $p$ is estimated by simulating
the transmission over a different BSC with a different cross-over
probability, denoted with $q$. Within this restriction the optimal
$q$, denoted $\hat{q}$, is the cross-over probability that minimizes
the IS estimator variance over all possible BSCs. Hereinafter, a general
formula for $\hat{q}$ is derived for any linear block code and for
any decoding algorithm.

Consider the parametric family of joint Bernoulli distributions $\mathcal{B}\left(wt\left(\mathbf{z}\right),q\right)$
generated by varying $q$ as IS distributions. The IS estimator for
WER in \eqref{eq:is-pe} specializes to
\begin{eqnarray}
\hat{P}_{IS}\left(e\right) & = & \frac{1}{N}\sum_{i=1}^{N}I\left(\mathbf{z}_{i}\right)\frac{f\left(\mathbf{z}_{i};p\right)}{f\left(\mathbf{z}_{i};q\right)}\\
 & = & \frac{1}{N}\sum_{i=1}^{N}I\left(\mathbf{z}_{i}\right)\frac{p^{wt(\mathbf{z}_{i})}\left(1-p\right)^{n-wt\left(\mathbf{z}_{i}\right)}}{q^{wt(\mathbf{z}_{i})}\left(1-q\right)^{n-wt\left(\mathbf{z}_{i}\right)}}\label{eq:IS-W-WER}\\
 & = & \frac{1}{N}\sum_{i=1}^{N}I\left(\mathbf{z}_{i}\right)W\left(wt(\mathbf{z}_{i});p,q\right),\label{eq:IS-WER-estimator}
\end{eqnarray}
where $\mathbf{z}_{i}\sim\mathcal{B}\left(wt\left(\mathbf{z}\right),q\right)$.
Under the assumption $\mathbf{c}=\mathbf{0}$, the estimator can be
equivalently expressed as
\begin{equation}
\hat{P}_{IS}\left(e\right)=\frac{1}{N}\sum_{i=1}^{N}I\left(\mathbf{e}_{i}\right)W\left(wt\left(\mathbf{e}_{i}\right);p,q\right),\label{eq:IS-Pe-estimator}
\end{equation}
where $\mathbf{e}_{i}\sim\mathcal{B}\left(wt\left(\mathbf{e}\right),q\right)$.
The general expression of the variance for the above estimator is
\begin{equation}
\mathrm{var}_{q}\left[\hat{P}_{IS}\left(e;q\right)\right]=\frac{\mathbb{E}_{q}\left[I\left(\mathbf{e}\right)W^{2}\left(wt\left(\mathbf{e}\right);p,q\right)\right]-P\left(e\right)^{2}}{N}\label{eq:VAR-IS}
\end{equation}
and clearly depends on $q$ through the weighting function $W\left(\cdot\right)$\cite{Smith1997}.
Therefore, the problem is to find the parameter $q$ that minimizes
(\ref{eq:VAR-IS}), i.e.
\begin{equation}
\hat{q}=\arg\min_{q}\mathrm{var}_{q}\left[\hat{P}_{IS}\left(e;q\right)\right].\label{eq:minVAR-problem}
\end{equation}

The expression of the IS estimator variance in the general case of
linear block codes is given by the following lemma.
\begin{lem}
\label{lem:IS-variance}The variance of $\hat{P}_{IS}\left(e;q\right)$
with importance sampling distribution in the parametric family $\mathcal{B}\left(i,q\right)$
is given by
\begin{equation}
\mathrm{var}_{q}\left[\hat{P}_{IS}\left(e;q\right)\right]=\frac{1}{N}\sum_{i=t+1}^{n}\left(W\left(i;p,q\right)P_{p}\left(e;i\right)-P_{p}\left(e;i\right)^{2}\right)\label{eq:IS-var}
\end{equation}
where $W\left(i;p,q\right)$ is the weighting function of the IS estimator;
$P_{p}\left(e;i\right)$ is the joint probability of decoding error
with $i$ errors over a BSC with cross-over probability $p$; $t$
is the error-correcting capability of the decoder.\end{lem}
\begin{IEEEproof}
The proof is given in Appendix \ref{sec:proof-of-lem-IS-variance}.
\end{IEEEproof}
The above lemma provides a general expression of the variance of the
IS estimator that depends on the specific decoding algorithm employed
\emph{only through the error-correcting capability of the decoder
$t$}. This parameter represents the maximum number of errors that
the decoder is able to correct and depends on the structure of the
linear block code and the decoding algorithm\cite{Benedetto1999}.

In order to solve the problem given by (\ref{eq:minVAR-problem})
we need to search for the equilibrium points of (\ref{eq:IS-var})
w.r.t. $q$. The following lemma gives a closed-form expression of
the variance derivative.
\begin{lem}
\label{lem:IS-variance-derivative}The derivative of the variance
of the IS estimator (\ref{eq:IS-Pe-estimator}) is given by
\begin{equation}
\frac{\partial}{\partial q}\mathrm{var}_{q}\left[\hat{P}_{IS}\left(e\right)\right]=-\frac{1}{N}\sum_{i=t+1}^{n}\frac{i-nq}{q\left(1-q\right)}W\left(i;p,q\right)P_{p}\left(e;i\right).\label{eq:IS-var-derivative}
\end{equation}
 \end{lem}
\begin{IEEEproof}
The proof is given in Appendix \ref{sec:proof-of-lem-IS-variance-derivative}.
\end{IEEEproof}
The solution of the minimization problem (\ref{eq:minVAR-problem})
can be obtained by equating to zero $\frac{\partial}{\partial q}\mathrm{var}_{q}\left[\hat{P}_{IS}\left(e\right)\right]$
if the IS variance is convex with respect to the variable $q$. The
following lemma states that the second derivative of the IS estimator
is always positive and then the variance of the IS estimator is convex. 
\begin{lem}
\label{lem:IS-convexity}The IS estimator (\ref{eq:IS-Pe-estimator})
is a convex function with respect to the variable $q$.\end{lem}
\begin{IEEEproof}
The proof is given in Appendix \ref{sec:proof-of-lem-IS-convexity}.
\end{IEEEproof}
The following theorem gives the general expression for the value of
$q$ that minimizes the variance of the IS estimator and for which
the estimation requires the minimum number of generated samples for
a fixed relative error.
\begin{thm}
\label{thm:min-var-q}The parameter $q$ that minimizes the variance
of the IS estimator given by (\ref{eq:IS-Pe-estimator}) is
\begin{equation}
\hat{q}=\frac{1}{n}\frac{\sum_{i=t+1}^{n}iW\left(i;p,\hat{q}\right)P_{p}\left(e;i\right)}{\sum_{i=t+1}^{n}W\left(i;p,\hat{q}\right)P_{p}\left(e;i\right)}.\label{eq:min-var-q}
\end{equation}
\end{thm}
\begin{IEEEproof}
The proof is obtained by solving $ $$\frac{\partial}{\partial q}\mathrm{var}_{q}\left[\hat{P}_{IS}\left(e\right)\right]=0$
and exploiting Lemma \ref{lem:IS-variance-derivative}.
\end{IEEEproof}
The result in \eqref{eq:min-var-q} defines implicitly the optimal
$q$ and therefore it is not possible to obtain a closed-form solution.
In some cases, however, (\ref{eq:min-var-q}) assumes a simplified
expression. When $np\ll1$ the following approximation holds \cite{Benedetto1999}

\begin{align}
\mathrm{var}_{q}\left[\hat{P}_{IS}\left(e\right)\right] & \simeq\frac{1}{N}W\left(t+1;p,q\right)P_{p}\left(e;t+1\right)\nonumber \\
 & -\frac{1}{N}P_{p}\left(e;t+1\right)^{2}\label{eq:approx-IS-estimator-variance}
\end{align}
and $\hat{q}$ can be expressed explicitly, as stated by the following
theorem.
\begin{thm}
\label{thm:min-var-pIS}Under the approximation $np\ll1$, the parameter
$q$ that minimizes the variance of the IS estimator (\ref{eq:IS-Pe-estimator})is
\begin{equation}
\hat{q}\simeq\frac{t+1}{n}.\label{eq:q-opt-approx}
\end{equation}
\end{thm}
\begin{IEEEproof}
The proof is given in Appendix \ref{sec:proof-of-thm-min-var-pIS}.
\end{IEEEproof}
A notable consequence of Theorem \ref{thm:min-var-pIS} is the independence
of $\hat{q}$ from the cross-over probability $p$ (which in turn
depends on the SNR), therefore leading to an SNR-invariant IS-MC simulation.
In this case estimation of WERs for a whole range of SNRs can be obtained
by running one IS-MC simulation with a BSC with parameter $\hat{q}$
given by \eqref{eq:q-opt-approx}, in the place of one simulation
for each SNR. Thus the whole performance curve WER versus SNR can
be obtained with a dramatic reduction of the number of samples to
be generated. It is also interesting to note that for the Hamming
code $\left(7,4\right)$ Eq. (\ref{eq:q-opt-approx}) gives $\hat{q}=2/7=0.2857$
which confirms the value of $\hat{q}$ that Sadowsky found empirically
in \cite{Sadowsky1990}. Furthermore, Sadowsky noted also the SNR
invariance of $\hat{q}$ with respect to $p$, without giving, unfortunately,
any explanation.

Note also that for short codes the assumption $np\ll1$ holds for
a large range of SNRs and then (\ref{eq:q-opt-approx}) is valid for
values of $p$ of interest, while for long codes the same assumption
holds only for high SNRs and (\ref{eq:q-opt-approx}) may not be useful
in practice. 

The result of Theorem \ref{thm:min-var-pIS} can also be used conversely
to estimate $t$ if an estimate $\hat{q}$ is available. In the next
section a method to estimate $\hat{q}$ is provided and then $t$.
This method is particularly useful for long codes when exhaustive
search for $d_{min}$ becomes computationally very intensive and/or
the decoder is not optimal and no explicit relationship between $d_{min}$
and $t$ is known.

\section{Algorithms}

\label{sec:algorithms}

The results presented in the previous section are exploited here to
formulate two different IS-MC simulation algorithms to obtain performance
curves in terms of WER vs SNR and an algorithm to estimate the error
correcting capability of the decoder. The two fast-simulation algorithms
compute the the WER estimate by means of the same IS estimator \eqref{eq:IS-Pe-estimator}
and they differ only in the choice of the Bernoulli IS distribution
parameter $q$. The first algorithm, called basic fast-simulation
algorithm (IS-MC basic), estimates the optimal value $\hat{q}$ and
then proceeds with WER estimation. It is the most general algorithm
since no specific assumption is required. The second algorithm assumes
$q=\left(t+1\right)/n$, a choice based on Th. \ref{thm:min-var-pIS},
and since $q$ is independent on the current SNR, the algorithm is
called SNR-invariant IS-MC algorithm. Under the assumption $np\ll1$
the SNR-invariant IS-MC algorithm is computationally more efficient
with respect to the IS-MC basic, as the same generated samples can
be used to estimate WERs at different SNRs. The choice between the
two algorithms depends on the code length $n$ and cross-over probability
$p$ (or, equivalently, the range of SNRs of interest) and therefore
on whether the assumption $np\ll1$ holds or not. 

Finally, the third algorithm is also based on the result of Th. \ref{thm:min-var-pIS}
and does not estimate the WER, but rather the error correcting capability
of the code.

\subsection{Basic fast-simulation algorithm (IS-MC basic)}

The basic version of the algorithm computes an estimate of the parameter
$\hat{q}$ iteratively, i.e. by updating $q$ at iteration $j$ from
the $q$ at iteration $j-1$. In fact, from \eqref{eq:min-var-q}
the following update rule can be derived
\begin{equation}
\hat{q}_{j}=\frac{1}{n}\frac{\sum_{i=t+1}^{n}iW\left(i;p,\hat{q}_{j-1}\right)P_{p}\left(e;i\right)}{\sum_{i=t+1}^{n}W\left(i;p,\hat{q}_{j-1}\right)P_{p}\left(e;i\right)},
\end{equation}
that can also be written as
\begin{equation}
\hat{q}_{j}=\frac{1}{n}\frac{\sum_{i=t+1}^{n}iW^{2}\left(i;p,\hat{q}_{j-1}\right)P_{q}\left(e;i\right)}{\sum_{i=t+1}^{n}W^{2}\left(i;p,\hat{q}_{j-1}\right)P_{q}\left(e;i\right)},\label{eq:update-q-1}
\end{equation}
since $P_{p}\left(e;i\right)=W\left(i;p,q\right)P_{q}\left(e;i\right)$.
Finally, the stochastic counterpart approximating \eqref{eq:update-q-1}
can be written in terms of the indicator function $I\left(\cdot\right)$
\begin{equation}
\hat{q}_{j}=\frac{1}{n}\frac{\sum_{i=1}^{N_{q}}I\left(\mathbf{z}_{i}\right)wt\left(\mathbf{z}_{i}\right)W^{2}\left(\mathbf{z}_{i};p,\hat{q}_{j-1}\right)}{\sum_{i=1}^{N_{q}}I\left(\mathbf{z}_{i}\right)W^{2}\left(\mathbf{z}_{i};p,\hat{q}_{j-1}\right)},\label{eq:update-q}
\end{equation}
where $\mathbf{z}_{i}\sim\mathcal{B}\left(wt(\mathbf{z}_{i}),\hat{q}_{j-1}\right)$.

In practice the IS simulation consists of two major steps. During
the first step an estimate of $\hat{q}$ is derived through (\ref{eq:update-q})
with a fixed number of iterations and in the second step the WER estimation
is performed by running the simulation with the IS Bernoulli distribution
with parameter $\hat{q}$ estimated in the first step. Even though
an additional step is required to derive $\hat{q}$, it is expected
that the total number of generated words will be reduced dramatically
w.r.t. the standard MC simulation given $\kappa$.

Algorithm \ref{alg:IS-Monte-Carlo} implements the basic algorithm.
The \texttt{while} loop implements the main part of the simulation
that stops when either the relative error \texttt{WERre} is less than
the given relative error \texttt{re} or the total number of generated
words is greater than \texttt{maxNumWords}. First iterations of the
algorithm compute the estimate of $\hat{q}$, with parameter $l$
controlling the number of iterations required. The number of words
$N$ is represented by the variable \texttt{minNumWordsIS}. After
$\hat{q}$ has been estimated then the algorithm starts estimating
the WER.

\begin{algorithm}
\caption{\label{alg:IS-Monte-Carlo}IS simulation with embedded estimation
of $\hat{q}$}
\texttt{%
\begin{algorithmic}
\STATE{totWords = 0}
\STATE{WERre = 1}
\STATE{$\hat{q}$ = $\hat{q}_{0}$}
\WHILE{ (WERre > re) and (totWords < maxNumWords)}
\STATE{ $\mathbf{z}$ = rand(n, numWords) < $\hat{q}$ } \COMMENT{ BSC output }
\STATE{ $\hat{\mathbf{m}}$ = decode($\mathbf{z}$) } \COMMENT{ decoder output }
\IF{ $\hat{q}$  has been estimated }
\STATE{ compute running estimate of the WER according to \eqref{eq:IS-WER-estimator} }
\STATE{ update totWords }
\IF{ totWords > minNumWords }
\STATE{ update relative error WERre } \COMMENT{ relative error }
\ENDIF
\ENDIF
\IF{ totWords < l*minNumWordsIS }
\STATE{ update $\hat{q}$ according \eqref{eq:update-q} }
\ENDIF
\ENDWHILE
\end{algorithmic}
}
\end{algorithm}

The search for $\hat{q}$ depends on the starting probability $\hat{q}_{0}$.
A bad choice of $\hat{q}_{0}$ may slow down the rate of convergence
of the estimation of $\hat{q}$ and after \texttt{l} iterations $\hat{q}_{l}$
might not be close to the optimal solution at all. It is important
to choose $\hat{q}_{0}$ such a way that important events can be generated
and a sufficiently number of errors are obtained to get an accurate
estimate of $\hat{q}$. Obviously, if $\hat{q}_{0}$ is close to the
optimal solution then a small number of iterations is required. If
the number of correctable errors $t$ is known then a possible choice
could be the $\hat{q}$ given by \eqref{eq:q-opt-approx}, even though
for $np\ll1$ a more computationally efficient simulation algorithm
is possible, as it will be shown in the next section. An alternative
choice can be made by observing that in a typical scenario the algorithm
is run to draw a performance curve as function of the SNRs or the
cross-over probability $p$. One can use the $\hat{q}$ estimated
with the simulation at the previous SNR as starting probability for
the current SNR, i.e. $\hat{q}_{0}\left(\mathrm{SNR}_{i}+\Delta\mathrm{SNR}\right)=\hat{q}_{l}\left(\mathrm{SNR}_{i}\right)$,
since for relatively small $\Delta\mathrm{SNR}$ the new optimal $\hat{q}$
is expected to be in the neighborhood of the previous $\hat{q}$.
Furthermore, at low SNRs the WER is usually high enough to require
a limited and acceptable number of generated samples even with standard
Monte-Carlo simulation and therefore at low SNRs the choice of $\hat{q}_{0}$
is less critical and can be chosen equal to the cross-over probability
$p$.

The structure of the algorithm is very similar in its formulation
to that presented in the context of cross-entropy method for simulation
of rare events in \cite{Rubinstein2004}. An application of the cross-entropy
method to the estimation of very low WERs of linear block codes has
been proposed in \cite{Romano2011}. The main difference with respect
to the algorithm proposed in this paper is that the WER estimator
in \cite{Romano2011} has been proven to minimize the cross-entropy
between the optimal IS solution and the parametric family of the joint
Bernoulli distributions. Differently, the estimator proposed in this
paper has minimum variance, thus \emph{leading to a different stochastic
update rule} for $\hat{q}$. The aforementioned update rule is proven
to converge since the IS estimator variance (within the Bernoulli
family) is convex and therefore one (global) minimum exists. Finally,
it is worth noticing that the two approaches lead to the same SNR-invariant
algorithm, as the result of Th. \ref{thm:min-var-q} holds in both
cases.

\subsection{SNR--invariant fast-simulation algorithm}

The result of the Theorem \ref{thm:min-var-pIS} suggests a more computationally
efficient IS-MC simulation algorithm that improves the basic algorithm
derived in the previous sub-section. In fact, under the assumptions
of the Theorem \ref{thm:min-var-pIS}, the $\hat{q}$ given by \eqref{eq:q-opt-approx}
does not depend on the current specific cross-over probability $p$
of the channel being simulated. Then, the same set of generated samples
with $\hat{q}$ can be used to calculate the estimate of the WER at
different SNRs. More specifically, in \eqref{eq:IS-WER-estimator}
the only term that depends on $p$ is the weight function $W\left(wt(\mathbf{z}_{i});p,q\right)$,
which is a deterministic function. Therefore given one set of $N$
realization of $\mathbf{z}_{i}\sim\mathcal{B}\left(wt(\mathbf{z}_{i});q\right)$
it is possible to compute the estimated WER for any $p$ for which
the approximation $np\ll1$ holds. In other words, with just one IS
simulation WERs for any SNR in the range of application of Theorem
\ref{thm:min-var-pIS} can be estimated. 

On the other hand, the estimated $\kappa$ that controls the number
of words to be generated depends on the current SNR. A conservative
rule for the choice of the relative error to be used in the stop condition
is to select the relative error corresponding to the highest SNR in
the given range, since, due to the monotonic decrease of the WER curve,
this guarantees that all the other relative errors will be smaller.

\subsection{Error-correcting capability estimation algorithm}

The first step of the basic algorithm can be used to estimate the
error correcting capability of the code and/or decoder, under the
assumption of relatively high SNR, as stated by Theorem \ref{thm:min-var-q}.
In fact, Eq. \eqref{eq:q-opt-approx} can be inverted to derive $t$
from $\hat{q}$, that can be estimated. Note that, since the solution
must be an integer, the estimate of $\hat{q}$ may not need to have
the same accuracy as that required for fast-simulation. Note also,
that especially for long codes, the number of generated words to obtain
$t$ is far less than the number of codewords.

\section{Examples}

\label{sec:examples}

\begin{figure}
\noindent \begin{centering}
\includegraphics[width=1\columnwidth]{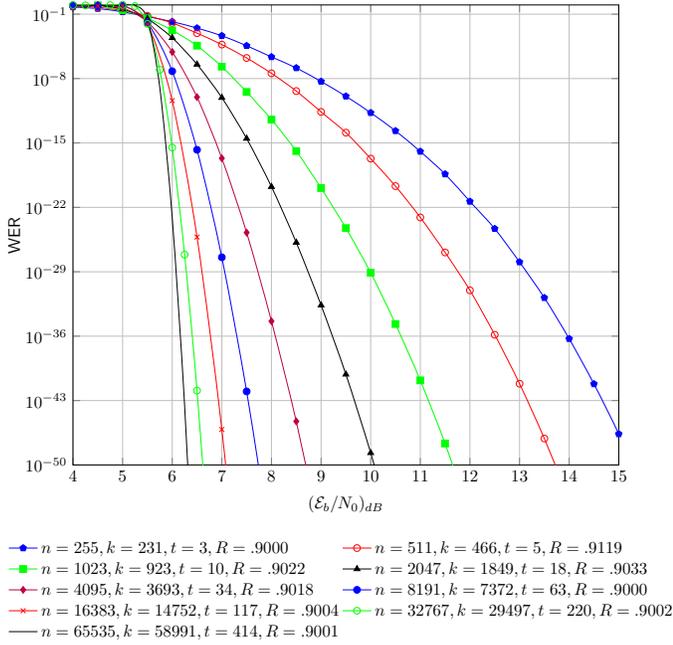}
\par\end{centering}

\caption{\label{fig:IS-MC}Estimated WER vs signal-to-noise ratio per uncoded
bit (in dB) with IS-MC basic fast-simulation algorithm for a set of
BCH codes with $R\simeq0.9$.}
\end{figure}

In this section some examples of applications of the proposed fast-simulation
algorithms are shown. The first example considers the application
of the IS-MC basic, by simulating performances of a set of BCH codes
\cite{Benedetto1999} with code rate $R=k/n\simeq0.9$, decoded with
the Berlekamp-Massey algorithm \cite{Wicker1995,Berlekamp1984}. In
Fig. \ref{fig:IS-MC} the WER vs signal-to-noise ratio per uncoded
bit in dB, $\left(\mathcal{E}_{b}/N_{0}\right)_{dB}$, is reported,
along with the parameters of the code that have been simulated. Each
curve is obtained by running the basic algorithm at different SNRs
with a stop condition on the relative error $\kappa=0.1$. For reliable
estimation of the parameter $\hat{q}$, the simulation of \texttt{minNumWords=$10^{2}$}
has been assured and only one iteration has been performed, i.e. \texttt{l}=1.
The results of each simulation run are plotted with points on the
interpolated curves, and correspond to the performances predicted
by the theoretical upper bound for linear block codes \cite{Benedetto1999}.
On the same set of BCH codes the error correcting capability estimation
algorithm has been applied with 100 generated words, and returns the
correct number of correctable errors.

\begin{figure}
\noindent \begin{centering}
\includegraphics[width=0.9\columnwidth]{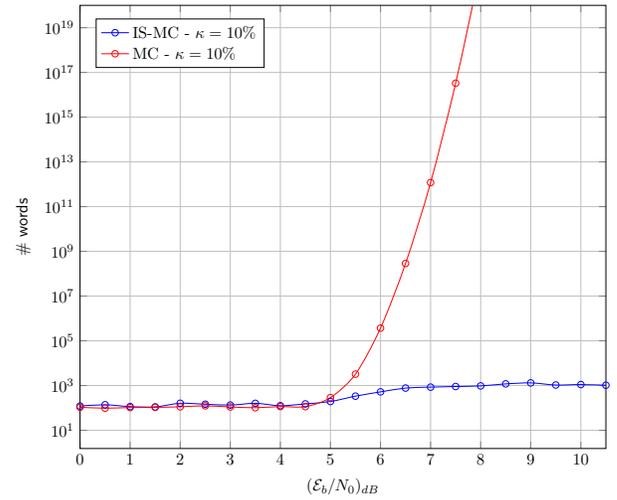}
\par\end{centering}

\caption{\label{fig:numWords-IS-MC-BCH}Number of generated words vs signal-to-noise
ratio per uncoded bit (in dB) for IS-MC basic and MC (estimated with
\eqref{eq:Nsamples}), BCH code $(2047,1849)$.}
\end{figure}

In Fig. \ref{fig:numWords-IS-MC-BCH} it is shown the number of generated
words required by a standard MC simulation with $\kappa=0.1$ for
BCH code $(2047,1849)$. The number, that includes also the number
of words required to estimate $\hat{q}$, increases with the SNR,
but at some point, in the IS case (blue curve), it reaches a steady
value. This corresponds to the region where the IS distribution does
not depend on the cross-over probability of the channel (cf. Th. \ref{thm:min-var-pIS}). 

\begin{figure}
\noindent \begin{centering}
\includegraphics[width=1\columnwidth]{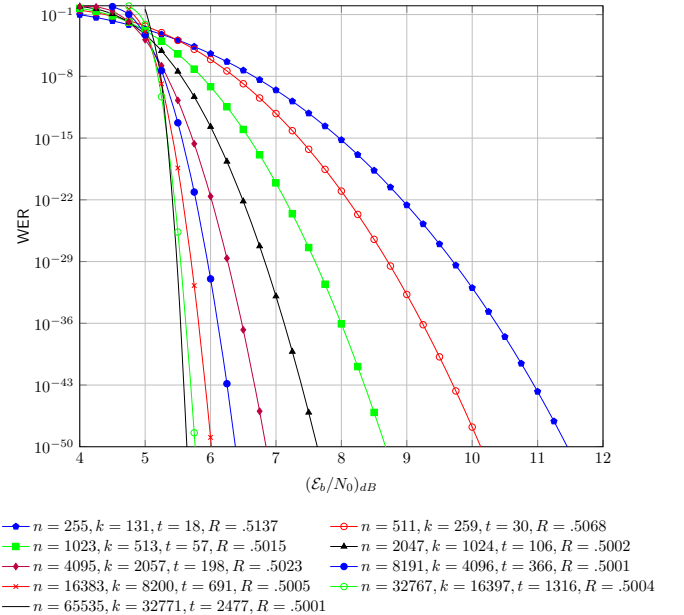}
\par\end{centering}

\caption{\label{fig:IS-MC-SNR-Invariant}IS-MC SNR-Invariant fast-simulation
algorithm for BCH code with $R\simeq0.5$.}
\end{figure}

A second example is shown in Fig. \ref{fig:IS-MC-SNR-Invariant} where
the performances in term of WER vs SNR for the SNR-invariant IS-MC
fast-simulation algorithm are plotted. In this case a different set
of BCH codes is considered, with a code rate $R\simeq0.5$. This set
presents a greater number of correctable errors, and thus the decoding
algorithm requires an increased computational complexity. The stop
condition has been set on the relative error estimated at the highest
SNR and only points with $\kappa<0.1$ has been plotted. The performances
in terms of WER confirm the theoretical results for BCH codes. More
interestingly, it is important to note that each curve has been obtained
with a single simulation run with a total number of generated words
reported in Tab. \ref{tab:words-IS-MC-SNR_invariant}: with approximately
$2\times10^{3}$ words it is possible to obtain the \emph{entire curve}
of performance. 

\begin{table}
\begin{centering}
\begin{tabular}{|c|c|}
\hline 
$\left(n,k\right)$ & \# of generated words\tabularnewline
\hline 
\hline 
$\left(255,231\right)$ & 980\tabularnewline
\hline 
$\left(511,259\right)$ & 1150\tabularnewline
\hline 
$\left(1023,513\right)$ & 1560\tabularnewline
\hline 
$\left(2047,1024\right)$ & 2030\tabularnewline
\hline 
$\left(4095,2057\right)$ & 2640\tabularnewline
\hline 
$\left(8191,7372\right)$ & 1710\tabularnewline
\hline 
$\left(16383,8200\right)$ & 2410\tabularnewline
\hline 
$\left(32767,29497\right)$ & 2040\tabularnewline
\hline 
$\left(65535,58991\right)$ & 2100\tabularnewline
\hline 
\end{tabular}
\par\end{centering}

\medskip{}

\caption{\label{tab:words-IS-MC-SNR_invariant}Number of generated words with
IS-MC SNR-invariant algorithm. For each BCH codes the total number
required to draw an entire performance curve is reported.}
\end{table}

\begin{figure}
\noindent \begin{centering}
\includegraphics[width=1\columnwidth]{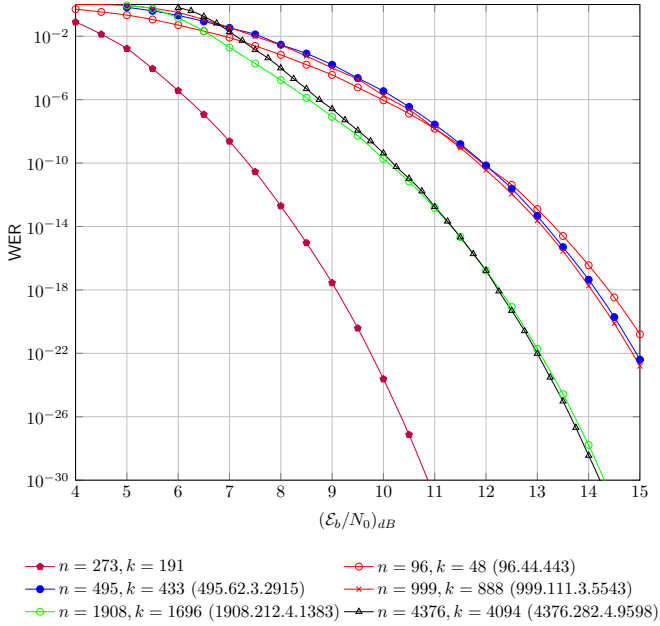}
\par\end{centering}

\caption{\label{fig:IS-MC-LDPC}IS-MC basic fast-simulation algorithm for a
set of LDPC codes. The code $\left(273,191\right)$ is taken from
\cite{Morelos-Zaragoza2006}, the others from \cite{mackay}.}
\end{figure}

The IS-MC method can be also employed to estimate the performances
of LDPC codes. Fig. \ref{fig:IS-MC-LDPC} shows the results of IS
simulations of a set of LDPC codes taken from \cite{mackay,Morelos-Zaragoza},
in terms of WER vs SNR per uncoded bit, for $\kappa=0.1$. All codes
are decoded with the bit-flip iterative algorithm described in \cite{Mahadevan2007},
with a number of iterations equal to 20. Estimation of $\hat{q}$
has been performed with $N=10^{3}$ generated words in one iteration.
The same number is the minimum enforced to obtain a reliable estimate
of the relative error. 

\begin{table*}
\begin{centering}
\begin{tabular}{|c|c|c|c|c|c|c|}
\hline 
code & $\left(96,48\right)$ & $\left(273,191\right)$ & $\left(495,433\right)$ & $\left(999,888\right)$ & $\left(1908,1696\right)$ & $\left(4376,4094\right)$\tabularnewline
\hline 
$t$ & 2 & 8 & 1 & 1 & 2 & $2$\tabularnewline
\hline 
\end{tabular}
\par\end{centering}

\medskip{}

\caption{\label{tab:number-of-correctable-errors}Estimated number of correctable
errors for LDPC codes of Fig. \eqref{fig:IS-MC-LDPC} with bit-flip
decoding \cite{Mahadevan2007}. }
\end{table*}

\begin{figure}
\begin{centering}
\includegraphics[width=1\columnwidth]{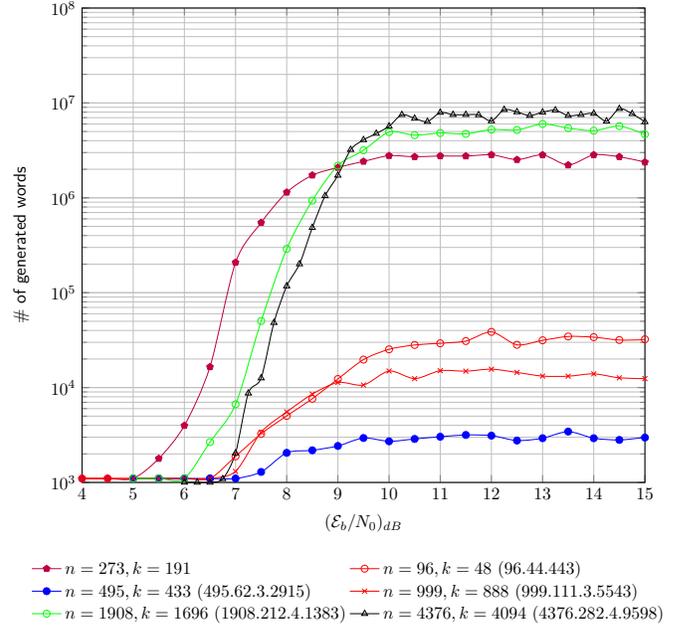}
\par\end{centering}

\caption{\label{fig:num-tot-words}Total number of generated words to obtain
results in Fig. \ref{fig:IS-MC-LDPC}.}
\end{figure}

The total number of generated words as function of the SNR is shown
in Fig. \ref{fig:num-tot-words}. It is interesting to note that,
as for BCH codes, at some point the number of generated words required
to achieve the prescribed relative error (i.e. $\kappa=0.1$) reaches
a steady value. The flat region reflects the independence of the IS
estimator variance on the SNR and identifies the SNR range over which
the SNR-invariant algorithm can be effectively applied. However, the
range of SNRs for which the curve is flat is different for each linear
block code as it depends on the IS estimator variance which in turn
depends on the structure of the code and the decoding algorithm. Numerical
results show also that the assumption $np\ll1$ is too strict, as
it would have as consequence a flat region starting at higher SNR
that those shown in Fig. \ref{fig:num-tot-words}. Furthermore, the
number of generated words in the flat region varies with the codes.
Results confirm the theoretical results obtained in Sec. \ref{sec:sub-optimal-importance-sampling}.

\begin{figure}
\noindent \begin{centering}
\includegraphics[width=1\columnwidth]{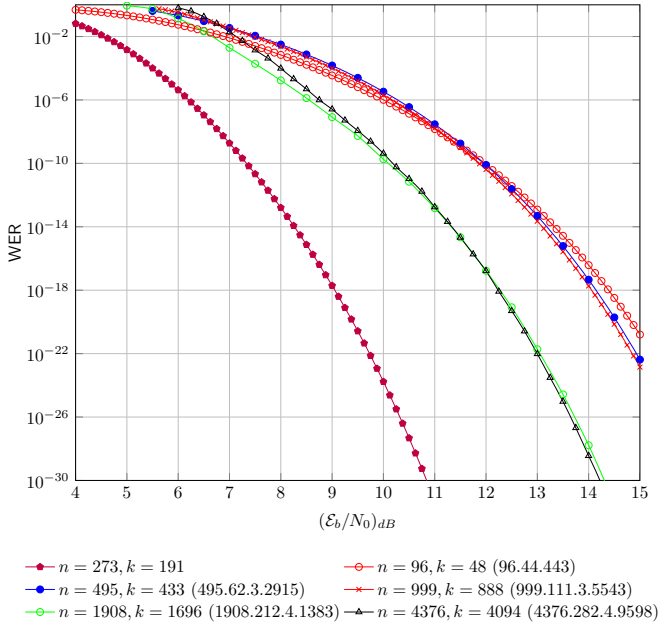}
\par\end{centering}

\caption{\label{fig:ldpc-snr-invariant}IS-MC SNR-invariant fast-simulation
algorithm for a set of LDPC codes with $\hat{q}=\left(t+1\right)/n$
and $t$ given by Table \ref{tab:number-of-correctable-errors}. The
code $\left(273,191\right)$ is taken from \cite{Morelos-Zaragoza2006},
the others from \cite{mackay}.}
\end{figure}

\begin{figure}
\noindent \begin{centering}
\includegraphics[width=1\columnwidth]{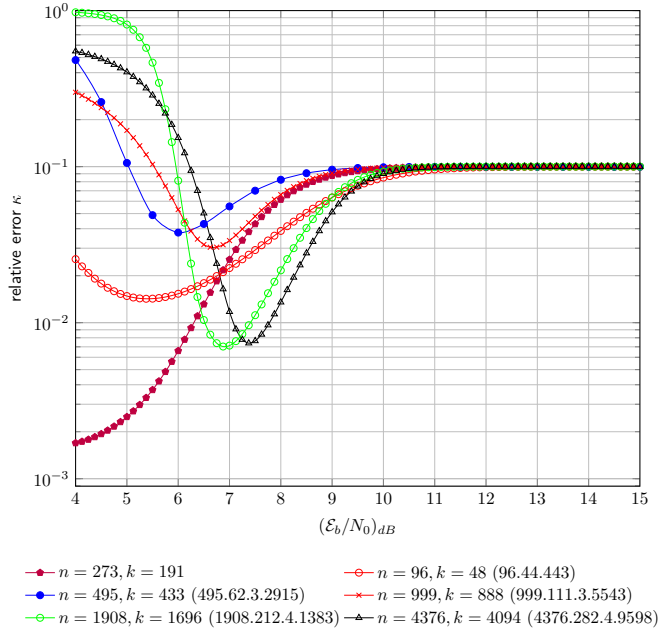}
\par\end{centering}

\caption{\label{fig:ldpc-snr-invariant-werre}Relative error as function of
$\left(\mathcal{E}_{b}/N_{0}\right)_{dB}$ corresponding to WER estimations
obtained by application of the SNR-invariant algorithm and reported
in Fig. \ref{fig:ldpc-snr-invariant}.}
\end{figure}

The error correcting estimation algorithm gives the number of correctable
errors shown in Tab. \ref{tab:number-of-correctable-errors}. Based
on these estimates, the IS-MC SNR-invariant method is employed to
draw the performance curves corresponding to the codes of Fig. \ref{fig:IS-MC-LDPC}.
Results are reported in Fig. \ref{fig:ldpc-snr-invariant}, that,
as expected, shows the same performance results as shown in Fig. \ref{fig:IS-MC-LDPC}.
The algorithm sets a stop condition on the relative error corresponding
to the WER estimate at the higher SNR (in this case $\mathcal{E}_{b}/N_{0}=15dB$)
and only WER estimates with relative error less than the given $\kappa=0.1$
are plotted. Results show also that the SNR-invariant algorithm correctly
estimates WER for a large range of SNRs. On the other hand, at very
low SNRs, the approximation \eqref{eq:min-var-q} becomes sensibly
different from the optimal solution. In Fig. \ref{fig:ldpc-snr-invariant-werre},
the relative error $\kappa$ vs $\mathcal{E}_{b}/N_{0}$ is plotted,
where becomes evident that \eqref{eq:min-var-q} at low SNRs is not
a good choice as the IS estimator variance increases up to a level
that makes the computational complexity of IS simulation even higher
that standard MC method or, equivalently, the relative error much
higher than the one obtained with the same number of generated words
with the standard MC method. Furthermore, it is interesting to note
that the range of SNRs for which the relative error is below $\kappa=0.1$
is larger than it was expected, suggesting that the assumption in
Th. \ref{thm:min-var-q} is too strict. 

\section{Conclusions}

\label{sec:conclusions}

In this paper an IS estimator for fast-simulation of linear block
codes with hard-decision decoding was presented. The estimator is
optimal, i.e. it has minimum variance, within the restriction of the
parametric family of IS distributions. It is possible to obtain huge
gains w.r.t. the standard MC in terms of generated words. Although
limited to the family of Bernoulli distributions, numerical examples
have shown that in most practical cases the gains obtained are significant.
However, the effective gain depends on the code and/or decoder performances
in terms of WER. The advantage of the proposed methods is the low
computational complexity and simplicity, since little modification
w.r.t. the standard MC simulation is required. Finally, higher gains
are achievable when the IS estimator does not depend on the cross-over
probability of the channel being simulated, typically at high SNR.

\section{Acknowledgments}

The authors would like to express their sincere gratitude to the Associate
Editor and the anonymous reviewers for taking their time into reviewing
this manuscript and providing comments that contributed to improve
the quality and the readability of the manuscript.

\appendices{}

\section{Proof of Lemma \ref{lem:IS-variance}}

\label{sec:proof-of-lem-IS-variance}The IS estimator can be rewritten
as weighted sum indexed by the weights of the error patterns
\begin{equation}
\hat{P}_{IS}\left(e\right)=\frac{1}{N}\sum_{i=t+1}^{n}N_{i}W_{i}
\end{equation}
where, for ease of notation, we denote $W\left(i;p,q\right)$ as $W_{i}$;
$N_{i}$ is the number of words with $i$ errors; $t$ is the maximum
number of errors that the decoder can correct; $N$ is the total number
of generated samples. Therefore the variance can be written as 
\begin{eqnarray}
\mathrm{var}_{q}\left[\hat{P}_{IS}\left(e\right)\right] & = & \mathrm{var}_{q}\left[\frac{1}{N}\sum_{i=t+1}^{n}N_{i}W_{i}\right]\\
 & = & \frac{1}{N^{2}}\sum_{i=t+1}^{n}\mathrm{var}_{q}\left[N_{i}W_{i}\right],
\end{eqnarray}
since generated samples constitute a realization of an i.i.d sequence
of random variables. The variance under the summation can be also
expressed as
\begin{eqnarray}
\mathrm{var}_{q}\left[N_{i}W_{i}\right] & = & \mathrm{var}_{q}\left[\sum_{j=1}^{N}I_{i}\left(\mathbf{z}_{j}\right)W\left(wt\left(\mathbf{z}_{j}\right);p,q\right)\right]\\
 & = & \mathrm{var}_{q}\left[\sum_{j=1}^{N}I_{i}\left(\mathbf{z}_{j}\right)W_{i}\right]\\
 & = & W_{i}^{2}\mathrm{var}_{q}\left[\sum_{j=1}^{N}I_{i}\left(\mathbf{z}_{j}\right)\right]\\
 & = & W_{i}^{2}\sum_{j=1}^{N}\mathrm{var}_{q}\left[I_{i}\left(\mathbf{z}_{j}\right)\right]
\end{eqnarray}
where $I_{i}\left(\cdot\right)$ is the indicator function that returns
1 when the event ``$\mathbf{z}_{j}$ contains $i$ errors'' occurs.
Note that the term $W_{i}$ is deterministic as it does not depend
on the random variable $\mathbf{z}_{j}$, $j=1,\dots,N$. Now define
\begin{equation}
P_{q}\left(e;i\right)\triangleq\sum_{\mathbf{z}}I_{i}\left(\mathbf{z}\right)f\left(\mathbf{z};q\right)
\end{equation}
as the joint probability that a decoding error occurs with an error
pattern of weight $i$, when the IS distribution is a Bernoulli with
parameter $q$. The variance of the estimator can be written as
\begin{eqnarray}
\mathrm{var}_{q}\left[N_{i}W_{i}\right] & = & W_{i}^{2}\sum_{j=1}^{N}P_{q}\left(e;i\right)\left(1-P_{q}\left(e;i\right)\right)\\
 & = & NW_{i}^{2}P_{q}\left(e;i\right)\left(1-P_{q}\left(e;i\right)\right)
\end{eqnarray}
The probability $P_{q}\left(e;i\right)$ can also be expressed in
terms of $P_{p}\left(e;i\right)$. By definition 
\begin{eqnarray}
P_{p}\left(e;i\right) & \triangleq & \sum_{\mathbf{z}}I_{i}\left(\mathbf{z}\right)f\left(\mathbf{z};p\right)\\
 & = & \sum_{\mathbf{z}}I_{i}\left(\mathbf{z}\right)\frac{f\left(\mathbf{z};p\right)}{f\left(\mathbf{z};q\right)}f\left(\mathbf{z};q\right)\\
 & = & W_{i}P_{q}\left(e;i\right)
\end{eqnarray}
Finally, the variance of the IS estimator is
\begin{multline}
\mathrm{var}\left[\hat{P}_{IS}\left(e\right)\right]=\frac{1}{N^{2}}\sum_{i=t+1}^{n}W_{i}^{2}NP_{q}\left(e;i\right)\left(1-P_{q}\left(e;i\right)\right)\\
=\frac{1}{N}\sum_{i=t+1}^{n}W_{i}^{2}P_{q}\left(e;i\right)\left(1-P_{q}\left(e;i\right)\right)\\
=\frac{1}{N}\sum_{i=t+1}^{n}W_{i}^{2}P_{q}\left(e;i\right)-\frac{1}{N}\sum_{i=t+1}^{n}W_{i}^{2}P_{q}\left(e;i\right)^{2}\\
=\frac{1}{N}\sum_{i=t+1}^{n}W_{i}P_{p}\left(e;i\right)-\frac{1}{N}\sum_{i=t+1}^{n}P_{p}\left(e;i\right)^{2}.
\end{multline}

\section{Proof of Lemma \ref{lem:IS-variance-derivative}}

\label{sec:proof-of-lem-IS-variance-derivative}The derivative of
$\mathrm{var}_{q}\left[\hat{P}_{IS}\left(e\right)\right]$ can be
written as
\begin{multline}
\frac{\partial}{\partial q}\mathrm{var}_{q}\left[\hat{P}_{IS}\left(e\right)\right]=\\
\frac{\partial}{\partial q}\left(\frac{1}{N}\sum_{i=t+1}^{n}\left(W\left(i;p,q\right)P_{p}\left(e;i\right)-P_{p}\left(e;i\right)^{2}\right)\right)\\
=\frac{1}{N}\sum_{i=t+1}^{n}\frac{\partial W(i;p,q)}{\partial q}P_{p}\left(e;i\right),\label{eq:derivative-VAR_q}
\end{multline}
where
\begin{equation}
P_{p}\left(e;i\right)\triangleq\sum_{\mathbf{z}}I_{i}\left(\mathbf{z}\right)f\left(\mathbf{z};p\right)
\end{equation}
does not depend on $q$. After some manipulations the derivative of
$W\left(i;p,q\right)$ w.r.t. $q$ can be written as 
\begin{multline}
\frac{\partial W\left(i;p,q\right)}{\partial q}=\frac{\partial}{\partial q}\left(\frac{p^{i}\left(1-p\right)^{n-i}}{q^{i}\left(1-q\right)^{n-i}}\right)\\
=-W\left(i;p,q\right)\left(\frac{i}{q}-\frac{n-i}{1-q}\right).\label{eq:derivative-W_i}
\end{multline}
By substituting (\ref{eq:derivative-W_i}) into (\ref{eq:derivative-VAR_q})
we obtain
\begin{multline}
\frac{\partial}{\partial q}\mathrm{var}_{q}\left[\hat{P}_{IS}\left(e\right)\right]=\\
=-\frac{1}{N}\sum_{i=t+1}^{n}W\left(i;p,q\right)\left(\frac{i}{q}-\frac{n-i}{1-q}\right)P_{p}\left(e;i\right)\\
=-\frac{1}{N}\sum_{k=t+1}^{n}\frac{i-nq}{q\left(1-q\right)}W\left(i;p,q\right)P_{p}\left(e;i\right).
\end{multline}

\section{Proof of Lemma \ref{lem:IS-convexity}}

\label{sec:proof-of-lem-IS-convexity}The convexity is proven by showing
that $\frac{\partial^{2}}{\partial q^{2}}\mathrm{var}_{q}\left[\hat{P}_{IS}\left(e\right)\right]>0$.
The second derivative of the variance is evaluated as follows (starting
from Eq. \eqref{eq:IS-var}) :
\begin{multline}
\frac{\partial^{2}}{\partial q^{2}}\mathrm{var}_{q}\left[\hat{P}_{IS}\left(e\right)\right]=\\
=\frac{\partial}{\partial q}\left\{ -\frac{1}{N}\sum_{i=t+1}^{n}\frac{i-nq}{q\left(1-q\right)}W\left(i;p,q\right)P_{p}\left(e;i\right)\right\} \\
=-\frac{1}{N}\sum_{i=t+1}^{n}P_{p}\left(e;i\right)\\
\left[\frac{\partial}{\partial q}\left(\frac{i-nq}{q\left(1-q\right)}\right)\cdot W\left(i;p,q\right)+\frac{i-nq}{q\left(1-q\right)}\cdot\frac{\partial W\left(i;p,q\right)}{\partial q}\right].\label{eq:second_der_var}
\end{multline}
After some manipulations, derivatives in (\ref{eq:second_der_var})
can be written as
\begin{eqnarray}
\frac{\partial}{\partial q}\left(\frac{i-nq}{q\left(1-q\right)}\right) & = & \frac{i\cdot(2q-1)-nq^{2}}{\left[q(1-q)\right]^{2}}\label{eq:firs_aux_derv}\\
\frac{\partial W\left(i;p,q\right)}{\partial q} & = & -W\left(i;p,q\right)\left(\frac{i}{q}-\frac{n-i}{1-q}\right)\label{eq:second_aux_der}\\
 & = & -W\left(i;p,q\right)\left(\frac{i-nq}{q(1-q)}\right)\label{eq:second_aux_der_2}
\end{eqnarray}
After plugging (\ref{eq:firs_aux_derv}) and (\ref{eq:second_aux_der_2})
into (\ref{eq:second_der_var}) the following expression is obtained
\begin{multline}
\frac{\partial^{2}}{\partial q^{2}}\mathrm{var}_{q}\left[\hat{P}_{IS}\left(e\right)\right]=\\
=-\frac{1}{N}\sum_{i=t+1}^{n}P_{p}\left(e;i\right)\\
\left[\frac{i\cdot(2q-1)-nq^{2}}{\left[q(1-q)\right]^{2}}\cdot W\left(i;p,q\right)-\left(\frac{i-nq}{q\left(1-q\right)}\right)^{2}\cdot W\left(i;p,q\right)\right]\\
=\frac{1}{N}\sum_{i=t+1}^{n}P_{p}\left(e;i\right)W\left(i;p,q\right)\left[\frac{\xi(q,i)}{q^{2}(1-q)^{2}}\right]
\end{multline}
where $\xi(q,i)\triangleq\left[(i-nq)^{2}+nq^{2}-i(2q-1)\right]$.
The sign of the second derivative depends only on the term $\xi(q,i)$
that can be rewritten as
\begin{eqnarray}
\xi(q,i) & = & i^{2}-2inq+n^{2}q^{2}+nq^{2}-2iq+i\\
 & = & n\left(1+n\right)q^{2}-2i\left(n+1\right)q+i\left(1+i\right)
\end{eqnarray}
The discriminant of the quadratic inequality $\xi\left(q,i\right)>0$
is given by
\begin{eqnarray}
\Delta & = & \left(-2i\left(n+1\right)\right)^{2}-4n\left(1+n\right)i\left(1+i\right)\\
 & = & 4i^{2}\left(n+1\right)^{2}-4ni\left(n+1\right)\left(i+1\right)\\
 & = & 4i\left(n+1\right)\left[i\left(n+1\right)-n\left(i+1\right)\right]\\
 & = & 4i\left(n+1\right)\left(ni+i-ni-n\right)\\
 & = & 4i\left(n+1\right)(i-n).
\end{eqnarray}
For $i<n$ we have $\Delta<0$, therefore the corresponding terms
in the sum that defines the second derivative are all positive. For
$i=n$ the term $\xi\left(q,n\right)$ is given by
\begin{eqnarray}
\xi\left(q,n\right) & = & \left(n-nq\right)^{2}+nq^{2}-n\left(2q-1\right)\\
 & = & n\left(1-q\right)^{2}+n\left(q^{2}-2q+1\right)\\
 & = & \left(n+1\right)\left(1-q\right)^{2}
\end{eqnarray}
which implies $\xi\left(q,n\right)\geq0$. The property $\xi\left(q,n\right)\geq0$
readily implies convexity of $\mathrm{var}_{q}\left[\hat{P}_{IS}\left(e\right)\right]$.

\section{Proof of Theorem \ref{thm:min-var-pIS}}

\label{sec:proof-of-thm-min-var-pIS}The best IS distribution in the
parametric family of Bernoulli distributions can be obtained by searching
the parameter $q$ that minimizes the variance of the IS estimator
(\ref{eq:IS-Pe-estimator}). From (\ref{eq:approx-IS-estimator-variance})
we have that the only term that depends on $q$ is $W\left(t+1;p,q\right)$,
denoted for convenience as $W_{t+1}$. In order to minimize the variance
of the IS estimator the term $W_{t+1}$ has to be to minimized, hence
\begin{equation}
\arg\min_{q}\mathrm{var}\left[\hat{P}_{IS}\left(e\right)\right]=\arg\min_{q}W_{t+1}
\end{equation}
or equivalently
\begin{multline}
\arg\min_{q}\mathrm{var}\left[\hat{P}_{IS}\left(e\right)\right]=\arg\min_{q}\ln W_{t+1}\\
=\arg\max_{q}\ln\left[q^{t+1}\left(1-q\right)^{n-t-1}\right].
\end{multline}
 The solution is obtained by equating the derivative of $\ln\left[q^{t+1}\left(1-q\right)^{n-t-1}\right]$
to zero and, after some manipulations, results to be
\begin{equation}
q=\frac{t+1}{n}
\end{equation}
The choice of $q$ according to the above equation minimizes the variance
of the IS estimator. Note that $q=0$ and $q=1$ cannot be solutions
of the minimization problem, as $t$ is always non negative and upper
bounded by $\left\lceil \left(d_{min}-1\right)/2\right\rceil $, where
$d_{min}$ is the minimum distance of the code that is always less
than $n$. From (\ref{eq:IS-var-derivative}) it is immediate to see
that for $q=0$ and $q=1$ the variance of the IS estimator presents
vertical asymptotes.

\bibliographystyle{ieeetr}
\bibliography{IEEEabrv,fsim_blockcodes}

\begin{IEEEbiography}[{\includegraphics[clip,width=1in,height=1.25in]{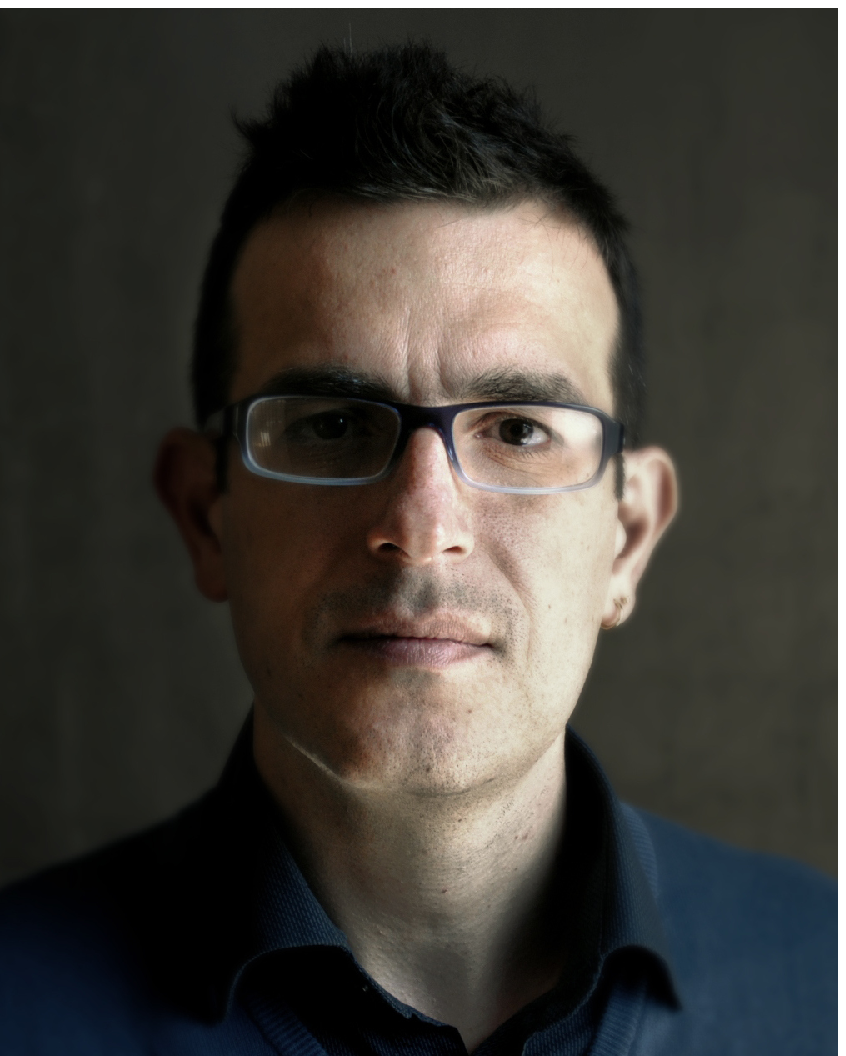}}]
{Gianmarco~Romano (M'11)} is currently Assistant Professor at
the Department of Information Engineering, Second University of Naples,
Aversa (CE), Italy. He received the ``Laurea'' degree in Electronic
Engineering from the University of Naples \textquotedblleft{}Federico
II\textquotedblright{} and the Ph.D. degree from the Second University
of Naples, in 2000 and 2004, respectively. From 2000 to 2002 he has
been Researcher at the National Laboratory for Multimedia Communications
(C.N.I.T.) in Naples, Italy. In 2003 he was Visiting Scholar at the
Department of Electrical and Electronic Engineering, University of
Conncticut, Storrs, USA. Since 2005 he has been with the Department
of Information Engineering, Second University of Naples and in 2006
has been appointed Assistant Professor. His research interests fall
within the areas of communications and signal processing. 
\end{IEEEbiography}

\begin{IEEEbiography}[{\includegraphics[clip,width=1in,height=1.25in]{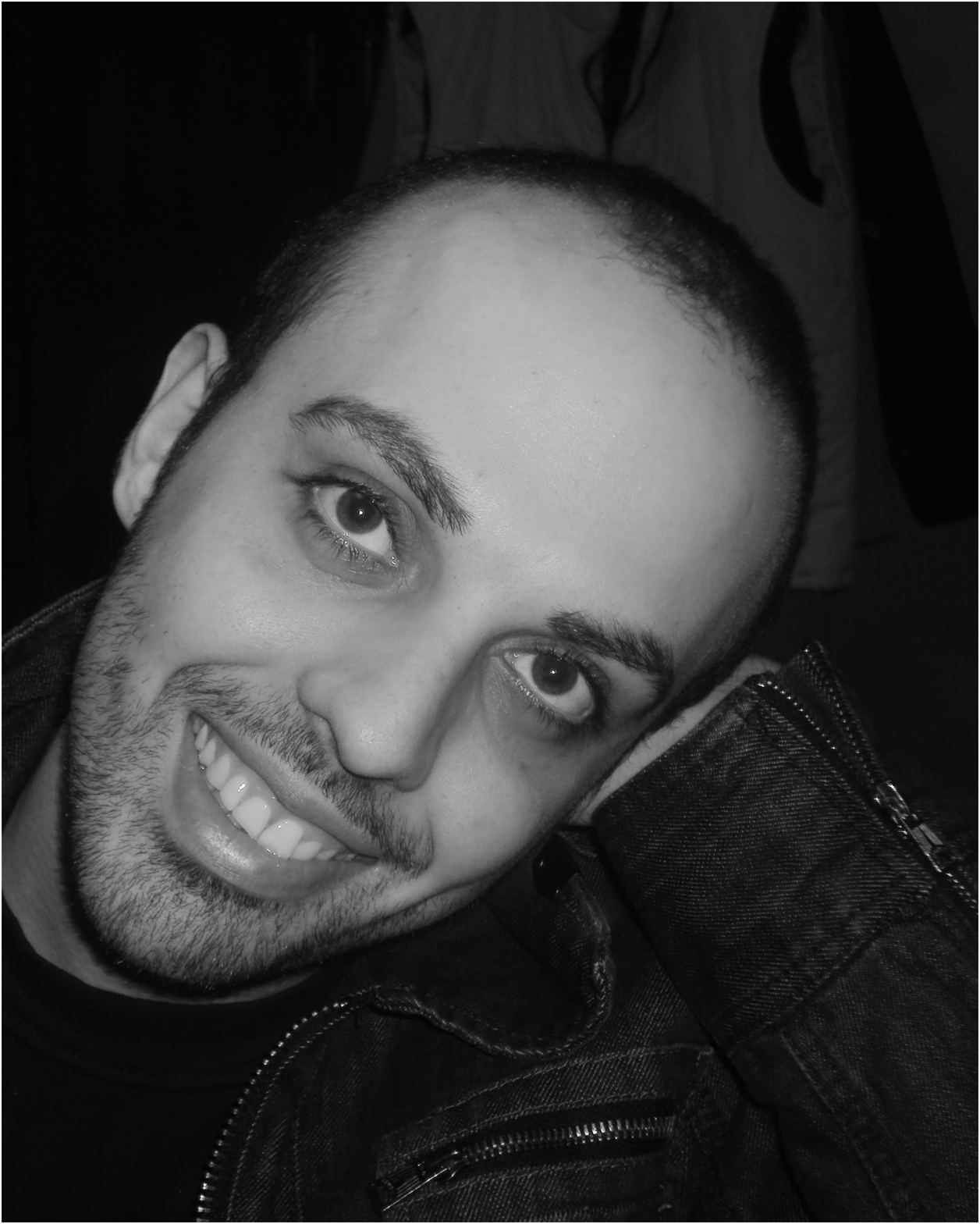}}]
{Domenico~Ciuonzo (S'11)}was born in Aversa (CE), Italy, on June
29th, 1985. He received the B.Sc. (\emph{summa cum laude}), the M.Sc.
(\emph{summa cum laude}) degrees in computer engineering and the Ph.D.
in electronic engineering, respectively in 2007, 2009 and 2013, from
the Second University of Naples, Aversa (CE), Italy. In 2011 he was
involved in the Visiting Researcher Programme of the former NATO Underwater
Research Center (now Centre for Maritime Research and Experimentation),
La Spezia, Italy; he worked in the \textquotedbl{}Maritime Situation
Awareness\textquotedbl{} project. In 2012 he was a visiting scholar
at the Electrical and Computer Engineering Department of University
of Connecticut (UConn), Storrs, US. He is currently a postdoc researcher
at Dept. of Industrial and Information Engineering of Second University
of Naples. His research interests are mainly in the areas of Data
and Decision Fusion, Statistical Signal Processing, Target Tracking
and Probabilistic Graphical Models. Dr. Ciuonzo is a reviewer for
several IEEE, Elsevier and Wiley journals in the areas of communications,
defense and signal processing\emph{.} He has also served as reviewer
and TPC member for several IEEE conferences.

\enlargethispage{-15cm}\end{IEEEbiography}

\end{document}